\documentclass[aps,pre,floats, twocolumn,showpacs,superscriptaddress,reprint]{revtex4}
\usepackage{graphicx}% Include figure files
\usepackage{graphics,dcolumn,bm,epic, eepic,fleqn,float}
\usepackage{amssymb,amsmath,amsfonts,multirow,rotate,color}
\usepackage{soul}
\usepackage{wasysym}
\usepackage{subfigure}

\definecolor{Myorange}{cmyk}{0,0.42,1,0}

\newcommand{\avg}[1]{\langle #1 \rangle}
\newcommand{\lay}[1]{^{[#1]}}

\begin{document}

\title{Irreducibility of multilayer network dynamics: the case of the
  voter model}

\author{Marina Diakonova}
\affiliation{Instituto de F\'isica Interdisciplinar y Sistemas Complejos IFISC (CSIC-UIB), E07122 Palma de Mallorca, Spain}
\author{Vincenzo Nicosia}
\affiliation{School of Mathematical Sciences, Queen Mary University of
London, E1 4NS Mile End Road, London, UK.}
\author{Vito Latora}
\affiliation{School of Mathematical Sciences, Queen Mary University of
London, E1 4NS Mile End Road, London, UK.}
\author{Maxi San Miguel}
\affiliation{Instituto de F\'isica Interdisciplinar y Sistemas Complejos IFISC (CSIC-UIB), E07122 Palma de Mallorca, Spain}

\begin{abstract}
We address the issue of the reducibility of the dynamics on a
multilayer network to an equivalent process on an aggregated
single-layer network. As a typical example of models for opinion
formation in social networks, we implement the voter model on a
two-layer multiplex network, and we study its dynamics as a function
of two control parameters, namely the fraction of edges simultaneously
existing in both layers of the network (edge overlap), and the
fraction of nodes participating in both layers (interlayer
connectivity or degree of multiplexity). We compute the asymptotic
value of the number of active links (interface density) in the
thermodynamic limit, and the time to reach an absorbing state for
finite systems, and we compare the numerical results with the
analytical predictions on equivalent single-layer networks obtained
through various possible aggregation procedures. We find a large
region of parameters where the interface density of large multiplexes
gives systematic deviations from that of the aggregates.
We show that neither of the standard aggregation procedures is able to
capture the highly nonlinear increase in the lifetime of a finite size
multiplex at small interlayer connectivity.  These results indicate
that multiplexity should be appropriately taken into account when
studying voter model dynamics, and that, in general, single-layer
approximations might be not accurate enough to properly understand
processes occurring on multiplex networks, since they might flatten
out relevant dynamical details.
\end{abstract}

\maketitle

\section{Introduction}
Real-world interactions often happen at different levels and are
therefore properly modelled by means of multilayer networks. Such
multilayer approaches \cite{Kivela2014,DeDomenico2013,Boccaletti2014}
have been applied to fields ranging from energy infrastructure
\cite{Buldyrev2010} and transport \cite{Morris2012,DeDomenico2013b,Kurant2006}, to epidemiology
\cite{Granell2013}.  The multilayer set up can either describe
interconnected networks with nodes of the same nature in each layer,
but interacting with nodes of different nature in a different layer,
or a multiplex structure with nodes of the same nature interacting via
a different network in each layer. In any case, a central
methodological question is that of multilayer reducibility, that is,
when the multilayer framework is really needed to explain new
phenomena, or when the system description can be reduced to an
appropriately aggregated or reformulated single-layer network. An
interesting contribution in this direction has recently come from the
study of the structural reducibility of multilayer networks, i.e. of
the possibility of aggregating some of the layers of a
multi-dimensional network while preserving its distinguishability from
the corresponding single-layer aggregated
graph~\cite{DeDomenico2015}. Although some recent works have pointed
out that multiplex dynamics can be intrinsically different from their
equivalent single-layer
counterparts~\cite{GomezGardenes2012,Sanz2014,Battiston2015RW,Battiston2015Ising,Chmiel2015},
little attention has been devoted so far to the problem of reducing a
process taking place on a multilayer network to a dynamically
equivalent process on an appropriate single-layer network. In this
paper we address this general question, by implementing the voter model 
on a multiplex networks and by studying the reducibility of its  
dynamics as a test case.  

The voter model~\cite{Liggett1975} is a nonequilibrium lattice
model~\cite{Redner2010} which gives a standard framework for studying
the influence of social imitation on the process of opinion formation
~\cite{Castellano2009}.  A basic question considered in this context
is when and how the system reaches an absorbing state with all the
interacting nodes in the same state, or when an active dynamical
situation of coexistence of different states prevails. The answer to
these questions is known to depend crucially on the network structure
and on the update rules
employed~\cite{Suchecki2005,Suchecki2005b,Redner2005,Redner2008,Juan2011}. The
voter model has also been instrumental to understand fundamental
phenomena in coevolution dynamics in which node states and network
structure have coupled dynamical evolution with two different time
scales~\cite{Vazquez2008,Gross2008,Gross2014,Diakonova2014,DiakonovaNoise}. In
terms of comparison with real data, a metapopulation voter model has
been recently shown to be able to account for voting patterns in the
US general elections~\cite{Jose2014}.

\begin{figure*}[!t]
  \begin{center}
    \includegraphics[width=6in]{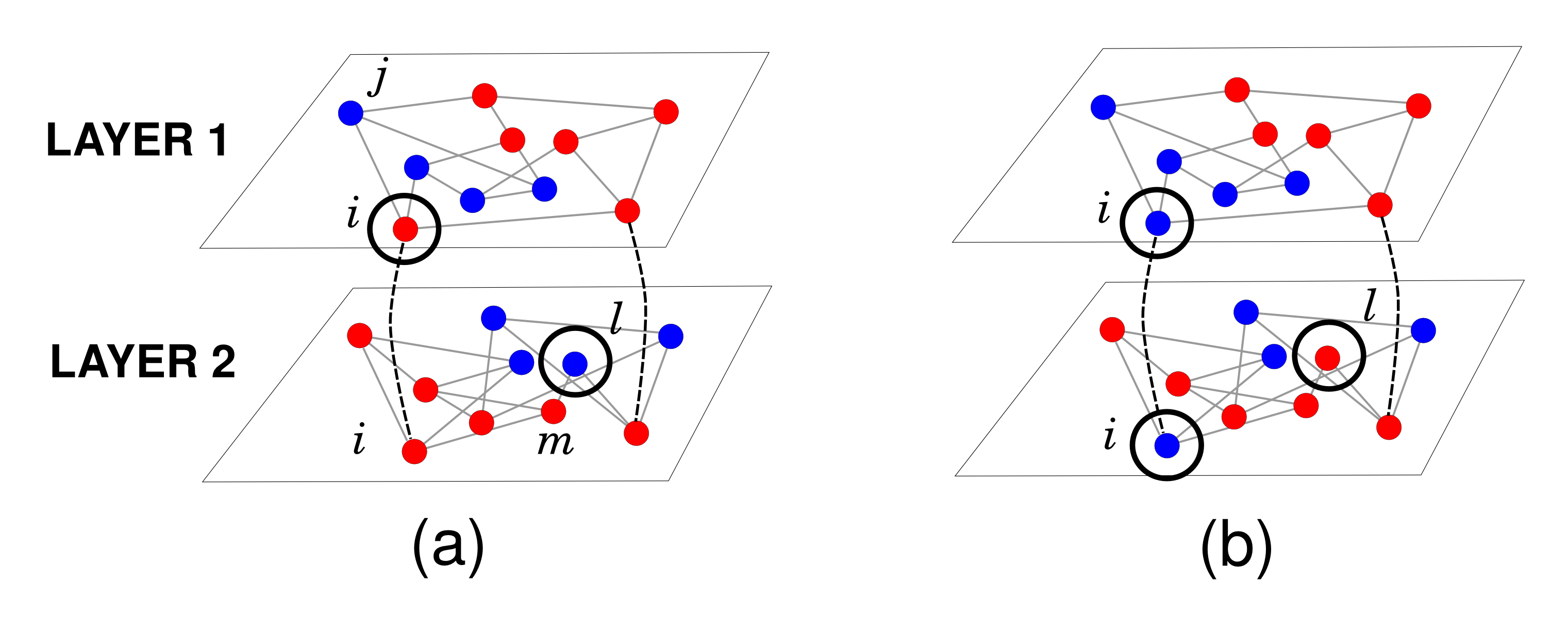}
  \end{center}
  \caption{(Color online) In the multiplex voter model each node $i$
    on layer $\alpha$ is associated to a state $s_{i}\lay{\alpha}(t)=
    \pm 1$ (the values $+1$ and $-1$ are respectively indicated in the
    figure by red and blue), which evolves according to one simple
    rule: select one of your neighbours $x$ on layer $\alpha $
    uniformly at random, and copy its state, i.e. set
    $s_{i}\lay{\alpha}(t+1) = s_{x}\lay{\alpha}(t)$. For instance, the
    state of node $l$ on layer $2$ changes from $s_{l}\lay{2}=-1$
    (blue, panel a) to $s_{l}\lay{2}=+1$ (red, panel b) since $l$
    selected its neighbour $m$ on layer $2$ and copied its state on
    that layer. The presence of inter-layer edges (dashed lines)
    denotes identification between nodes at the two layers. An example
    is the inter-layer edge connecting the two replicas of node
    $i$. In this case, any change in the state of node $i$ at one
    layer will enforce a change of its state on the other layer. In
    the figure, the state of node $i$ flips from red (panel a) to blue
    (panel b) on both layers, as a consequence of the interaction of
    node $i$ with its neighbour $j$ on layer $1$.}
  \label{fig:fig0}
\end{figure*}

Quite frequently, social interactions happen on different concurrent
contexts, so that any model of social imitation based on a
single-layer representation of social relationships should probably be
regarded only as a first-order approximation of a potentially more
complex dynamics. In order to better mimic the multidimensional nature
of social interactions, we consider here a multiplex voter model,
where agents interact at two distinct layers. We assume also that the
system exhibits a certain level of multiplexity, meaning that a
fraction $q$ of the agents is present on both
layers~\cite{Diakonova2014}. As a consequence, any change in the state
of those agents on either of the two layers is automatically
transferred to the other layer, effectively coupling the voter model
dynamics taking place on each of the two networks. By taking into
account heterogeneity in the participation of agents to layers we aim
at reproducing an interesting feature of real-world multilevel social
systems~\cite{Nicosia2014corr}. In particular, it has been shown that
the percentage of users of an online social network (e.g., Twitter)
which has also an account on another online social network (e.g.,
Facebook or LinkedIn) lies somewhere between $30\%$ and $70\%$, as
reported by the social media matrix periodically published by the Pew
Research Center~\cite{pewinternet}. These empirical findings confirm
the necessity to consider systems whose multiplexity is neither 0 nor
1, but somewhere in between.
A second important feature of real-world social interactions is the
fact that networks corresponding to different contexts share a number
of common links, as can be found for instance in data from online
games: analysis of the PARDUS online society
in~\cite{Szell2010,Klimek2013} yields significant overlap between
designation of other players as friends, correspondents, or trading
partners. This property of multilayer networks can be quantified by
the edge-overlap parameter~\cite{Battiston2014}.

In this work we address a fundamental question, that is whether the
coupling of several voter models into a multiplex dynamics, motivated
by the multiplicity of contexts influencing real interactions among
individuals, gives rise to qualitatively different phenomena or
effects than those observed in the classical single-layer setup. In
particular, we want to establish whether multiplexity makes any
difference for a simple social dynamics like the voter model, or if
instead the multiplex voter model can indeed be reduced to an
equivalent voter model dynamics on an appropriately constructed
single-layer network.

A preliminary result on the question of the multilayer reducibility
exists for a bilayer of uncorrelated networks evolving according to
the voter model with adaptive links~\cite{Diakonova2014}, but in the
limit of zero edge overlap. In this model each layer is associated
with a network plasticity parameter, that controls the rate at which
relations among agents are rearranged, such that if the values of
plasticity at the two layers are sufficiently different the system
displays a network \emph{shattered} fragmentation. On the other hand
when the layers have the same plasticity one finds that the results
coming from a pair approximation in the thermodynamic limit are
equivalent to those of an appropriate aggregated single-layer network 
\cite{Diakonova2014}. We build upon this, in time independent
networks, examining both the asymptotic properties of the
thermodynamic limit and the characteristic times to reach an absorbing
state for finite systems. We also consider different schemes to obtain
a possible equivalent aggregated single-layer network and highlight the
importance of the degree of multiplexity and edge overlapping.

The paper is organized as follows. Sect II introduces our voter
dynamics multilayer model. Sect. III describes our numerical
findings. These are compared in Sect. IV with single-layer theoretical
results for two natural methods of collapsing the multilayer into an
aggregated single-layer network. Sect. V analyzes an optimum single-layer
reduction method. Conclusions are discussed in Sect. VI.

\section{The Model} % (fold)
\label{sec:the_model}
We consider here the case of a 2-layer undirected unweighted multiplex
network, described by the pair of binary adjacency matrices
$\{A\lay{1}, A\lay{2}\}$, where $A\lay{\alpha}\equiv
\{a\lay{\alpha}_{ij}\}$, and $a\lay{\alpha}_{ij}=1$ if and only if
node $i$ and node $j$ are connected by a link at layer $\alpha$, and zero otherwise. On
each layer we have $N$ nodes. A parameter of interest in this study is
the {\em average edge overlap} $\omega$, that is the probability that an
edge is present on both layers:
\begin{equation}
  \omega = \frac{\sum_{i,j}a\lay{1}_{ij}a\lay{2}_{ij}}{2K}
\end{equation}
where
\begin{equation*}
  K=\sum_{i}\sum_{j>i}\frac{a\lay{1}_{ij}+a\lay{2}_{ij}}
  {1+a\lay{1}_{ij}a\lay{2}_{ij}}
\end{equation*}
is the number of edges of the graph obtained by aggregating the two
layers into a single one~\cite{Battiston2014}. Notice that $\omega=0$
only if each edge exists in exactly one of the two layers, but not in
the other one, while we have $\omega=1$ only if all the edges exist on
both layers.

Each node $i$ on layer $\alpha$ is associated to a binary state
variable $s\lay{\alpha}_i(t)$, where $s\lay{\alpha}_i(t)$ can be
either $+1$ or $-1$. 
Moreover, we assume that a fraction $q$ of the $N$ nodes participates
in both layers, requiring that if $i$ is one of these $qN$ nodes then
its state at the two layers will be identical at the end of every
update. Such nodes participating in both layers are chosen randomly at
initialization.
We can think of the parameter $q$ as the {\em interlayer connectivity} 
or the {\em degree of structural multiplexity} of the system. 
The model is illustrated in Fig.~\ref{fig:fig0}.

The multiplex voter dynamics consists of
a sequence of time steps. During a time step we perform $N$ updates,
and each update consists of three elementary operations, as follows:
\textit{i)} a layer $\alpha$ is selected at random, with uniform
probability; \textit{ii)} one of the nodes $i$ on layer $\alpha$ is
chosen at random and its state $s\lay{\alpha}_i$ is updated according
to the classical voter model dynamics, that is, $s\lay{\alpha}_i$
becomes the same as that of a randomly chosen neighbour of $i$ on
layer $\alpha$; \textit{iii)} if the updated node $i$ participates in
both layers, then the state of the corresponding node in the other
layer $\beta$ changes as well by setting
$s\lay{\beta}_i=s\lay{\alpha}_i$.  With this third operation, state
changes can propagate across layers: indeed the presence of a fraction of
nodes existing in both layers intertwines the voter
dynamics on the two layers, so that in general the evolution of the
overall multiplex dynamics might differ from the one we would observe
on two independent networks of the same size. As a limiting case, the
dynamics reduces to that of a classical voter model on a single-layer 
network only when $q=1$.

It is well known that in connected finite-size single-layer networks
the voter model dynamics always reaches an absorbing state, where all
the nodes have exactly the same state, in a survival time that scales
with the system size $N$~\cite{Suchecki2005, Redner2005, Redner2008,
  Castellano2009, Redner2010}.  In networks of high effective
dimensionality (including random networks), when $N\to\infty$ the
dynamics sustains an active disordered state in which nodes continue
to change their state~\cite{Suchecki2005b}. Such active state is the
one observed asymptotically in large systems, before finite-size
fluctuations pull the system towards the absorbing state. The
classical order parameter to measure the activity in the voter model
is the so-called {\em interface density} $\rho(t)$, defined as the
fraction of active edges of the network, i.e. of those edges whose
endpoints have different states.  In the multiplex model we can
define:
\begin{equation}
  \rho\lay{\alpha}(t) = \frac{1}{2K\lay{\alpha}}
  \sum_{i}\sum_{j<i}a\lay{\alpha}_{ij}|s\lay{\alpha}_i(t) -
  s\lay{\alpha}_j(t)|,
\end{equation}
for each layer $\alpha$, $\alpha = 1,2$, where we denote by
$K\lay{\alpha}=\frac{1}{2}\sum_{i,j}a\lay{\alpha}_{ij}$ the total
number of edges at layer $\alpha$. Notice that $\rho=0$ if and only if
all the nodes have the same state, while larger values of $\rho$ are
associated to active configurations. A second quantity of interest for
a finite system is the average time $\left< T \right>$ to reach an
absorbing state of consensus. Such a time can be defined as $\left< T
\right> = \int_0^{\infty} P_s(t)dt$ from the survival probability
$P_s(t)$, i.e. the probability for a system to be active at time
$t$~\cite{Juan2013}.

For single-layer uncorrelated networks at sufficiently large times $t
> N$, $P_s(t)$ decreases exponentially $P_s(t) \sim e^{-2t/\tau}$
\cite{Vazquez2008NJP}. Hence, $\left< T \right> \sim \tau/2$, so that
the dependencies of $\left< T \right>$ on system size and on the
moments of the degree distribution are given precisely by the
corresponding dependencies of the {\em characteristic time}
$\tau$. For uncorrelated networks one can find exact expressions for
the value of the average interface density in the thermodynamic limit
$\rho^{\rm single}$, as well as for the characteristic time to reach
the absorbing state $\tau^{\rm single}$
\cite{Vazquez2008,Vazquez2008NJP}. An important feature is that these
analytical predictions of $\rho^{\rm single}$ and $\tau^{\rm single}$
depend only on the values of the first two moments of the degree
distribution $P(k)$ of the network, i.e. on $\mu_1 = \avg{k} = \sum_k
k P(k) $ and $\mu_{2} = \avg{k^2}= \sum_k k^2 P(k)$, and not on any
other microscopic property of the network, and read:
\begin{equation}
  \rho^{\rm single} = \frac{\mu_1 - 2}{3(\mu_1 - 1)},
  \label{eq:rho_single}
\end{equation}
where $\rho^{\rm single}$ is the average over surviving runs, and
\begin{equation}
  \tau^{\rm single} = \frac{(\mu_1 -1)\mu_1^2 N}{(\mu_1-2)\mu_2}.
  \label{eq:tau_single}
\end{equation}
In the following we will focus on the values $\rho(q, \omega)$ and
$\tau(q, \omega)$ of interface density and characteristic time of the
multiplex voter model as a function of the average edge overlap of the
system $\omega$ and of the fraction $q$ of nodes present in both
layers.

\begin{figure*}%[!t]
  \begin{center}
    \subfigure[Interface density $\rho$]{
      \includegraphics[width=2in]{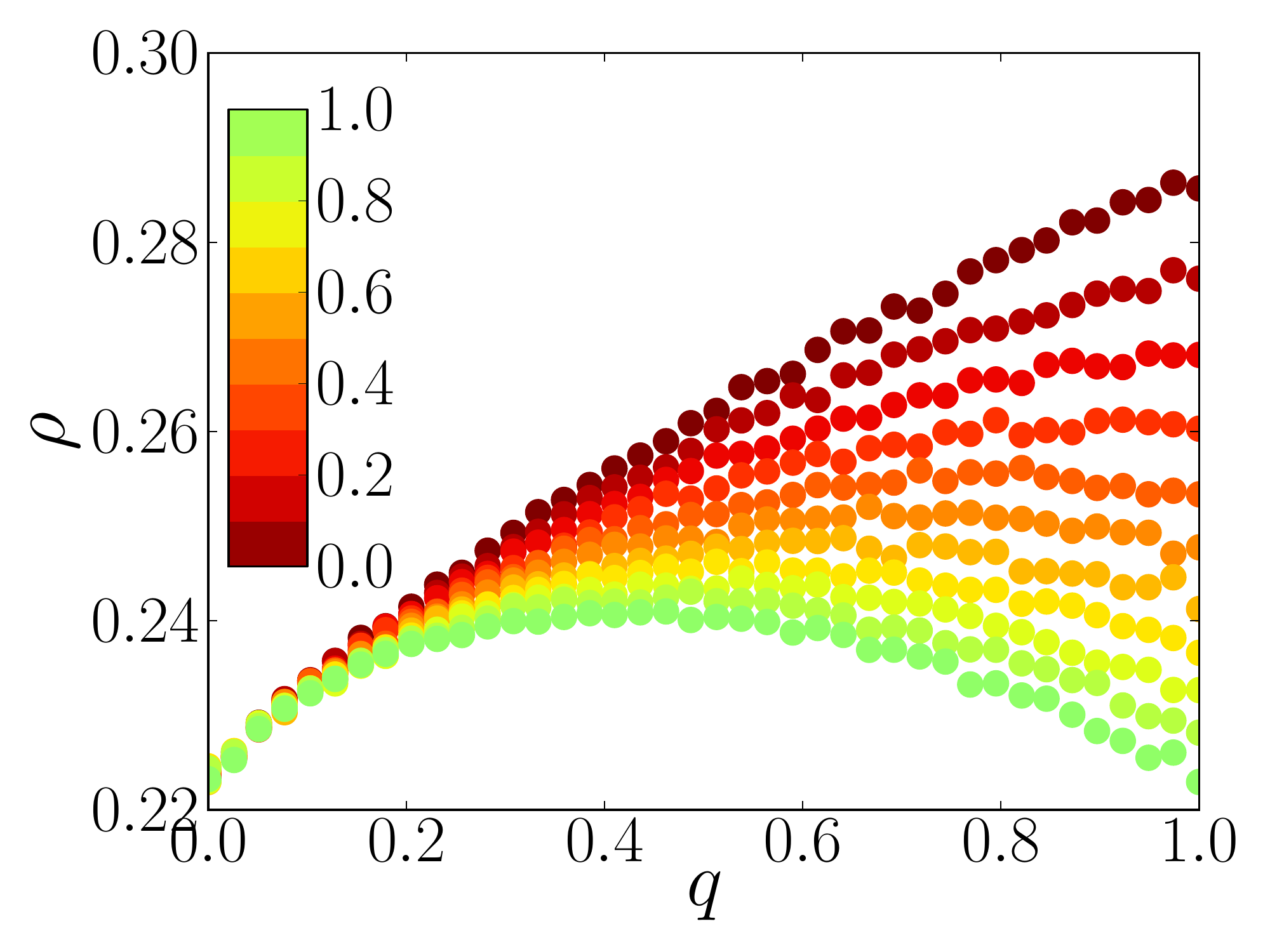}
	  \label{fig:numericalresults:a}
	}
    \subfigure[Survival Probability $P_s(t)$]{
      \includegraphics[width=2in]{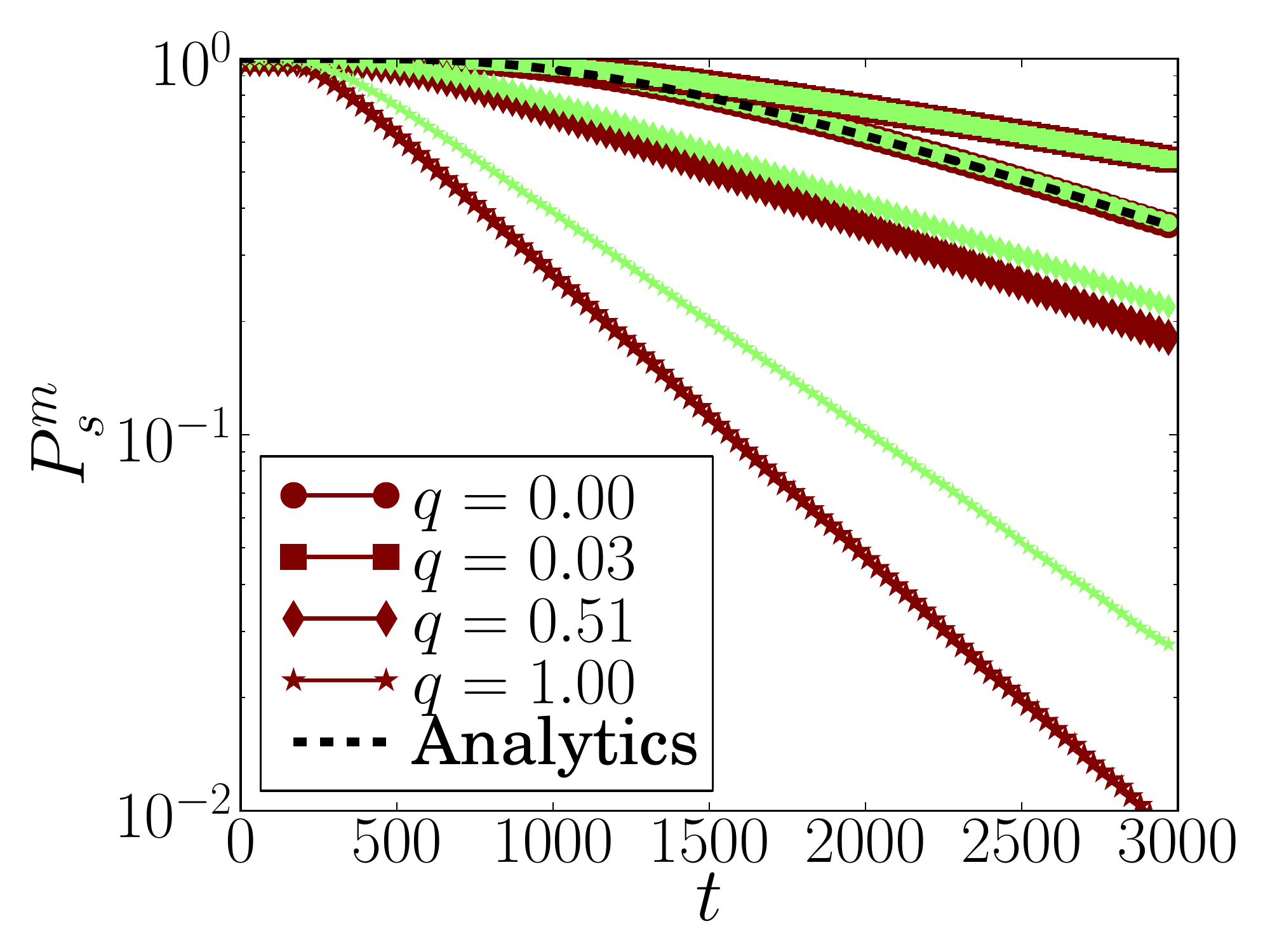}
	\label{fig:numericalresults:b}
	}
	\subfigure[Rescaled characteristic time $\tau/N$]{
      \includegraphics[width=2in]{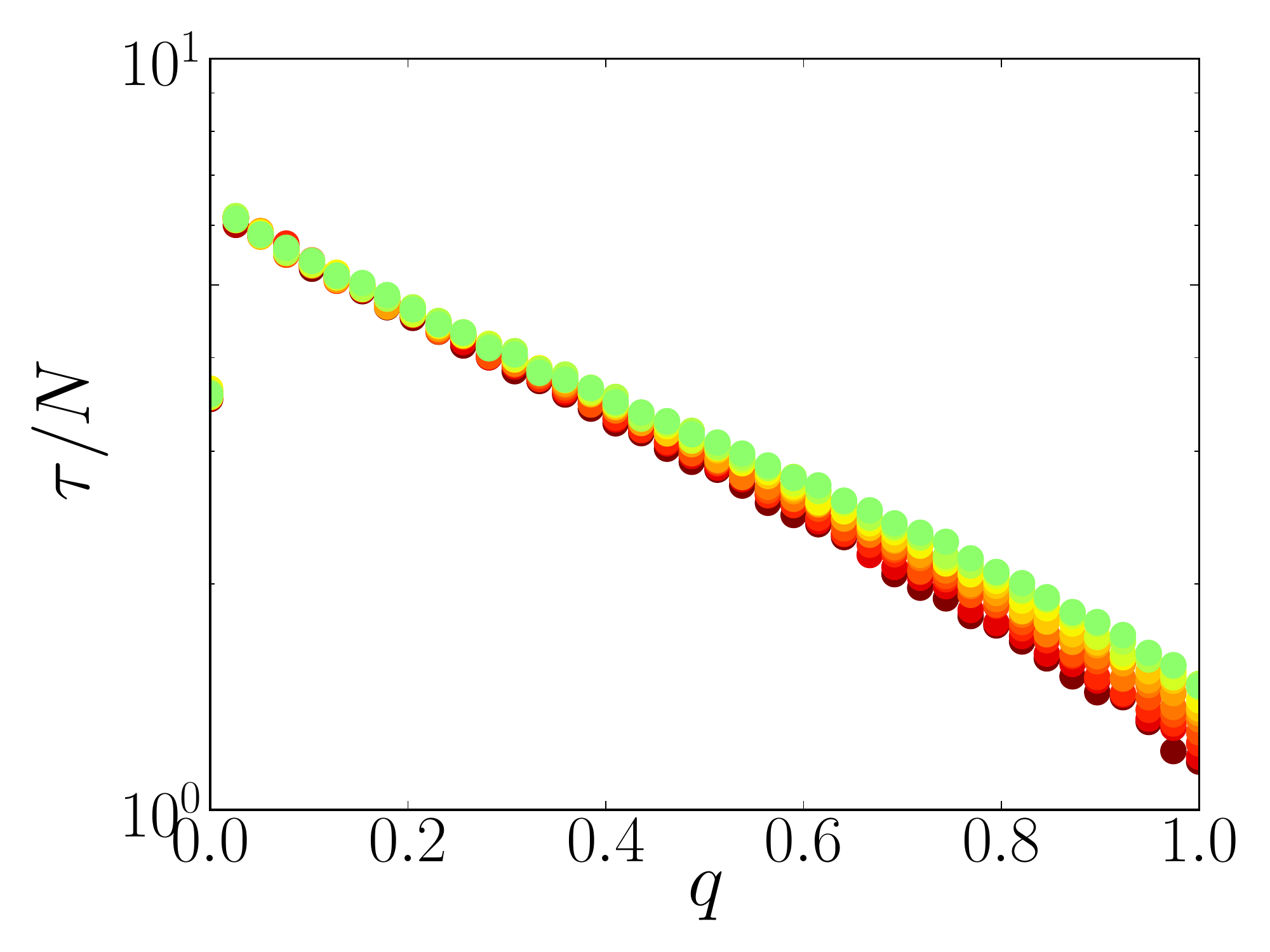}
	\label{fig:numericalresults:c}
	}
  \end{center}
  \caption{(Color online) (a) Asymptotic value of the interface
    density averaged over the surviving runs; (b) Survival probability
    of the multiplex for different $q$ values, for $\omega = 0$
    (maroon, larger and darker markers), and $\omega = 1$ (green,
    smaller and lighter markers). The line denoted as \emph{analytics}
    is $P_s^m(t,q = 0, \omega = 0)$ whose expression is given in the
    main text; (c) Logarithm of the average characteristic time $\tau$
    of the approach to the absorbing state, rescaled by $(1/N)$
    (legend same as in fig. \ref{fig:numericalresults:a}).  Quantities
    are functions of multiplexity $q$, and edge overlap $\omega$.  We
    have considered an ensemble of $10^5$ random initializations on a
    fixed duplex whose layers are random regular graphs, each with $N
    = 1000$ and $\avg{k}= 4$.}
  \label{fig:numericalresults}
\end{figure*}

\section{Numerical results}

We studied the voter model on a duplex network made by two random
regular graphs with $N$ nodes, each node having a degree equal to
$\mu$. We therefore have $\mu_1 = \avg{k\lay{1}}=\avg{k\lay{2}} =
\mu$, and $\mu_2 = \mu^2$. In our simulations we fixed $N=1000$ and
$\mu=4$, and we studied the dynamics of the system by varying the
value of the edge overlap $\omega$ and the fraction $q$ of nodes
participating in both layers. Since the system is finite, it will
eventually converge to the absorbing frozen state corresponding to an
interface density equal to zero. Consequently, for each time $t$, we
will evaluate the average value of the interface density only over the
surviving runs, i.e. on those realisations of the dynamics which are
still active at time $t$. We associate the asymptotic value of this
numerically-obtained quantity with the value at the thermodynamic
limit, given by Eq.~(\ref{eq:rho_single}), and refer to it as
interface density, while being clear whether we are referring to the
numerical or analytical results. In the case of a network with two
layers, we have of course two values of interface density, one for
each layer. If $\avg{k\lay{1}} = \avg{k\lay{2}}$ the ensemble averages
on the two layers will be equal, and hence the activity of an
arbitrary layer is representative of the typical activity of the
entire system.  We therefore use the asymptotic value of the interface
density of an arbitrary layer to reflect the activity of the
multiplex, and dispense with the $\alpha$ superscript.

In Fig.~\ref{fig:numericalresults:a} we report the values of interface
density $\rho(q, \omega)$ as a function of $q$, and for different
values of $\omega$.  Let us consider first the case $\omega = 0$ of no
overlap between the structure of the two layers of the network. When
$q=0$, i.e. when inter-layer state passing is not allowed, the system
effectively corresponds to two identical but independent single-layer
voter dynamics, so that $\rho(q, \omega)$ is in accordance with the
classical analytical predictions for single-layer networks (since
$\mu_1 = 4$, Eq.~(\ref{eq:rho_single}) gives the activity $\rho(0,0) =
0.22$).  On the other hand, when $q=1$, i.e. when all the nodes
participate in both layers and each edge exists only in one layer, the
system is in all respects identical to a single-layer network with
$\mu_1 = 2 \mu$ (for $\mu_1 = 8$, Eq.~(\ref{eq:rho_single}) gives the
activity $\rho(1,0) = 0.286$).  For intermediate values of $q$ the
dynamics interpolates monotonically between the two extreme cases,
i.e. two dynamically indistinguishable voter models on single-layer
networks with $\mu_1 = \mu$ ($q=0$), and one voter model on a
single-layer network with $\mu_1 = 2 \mu$ ($q=1$).  The picture
changes completely as soon as $\omega$ is large enough.  In general,
when the overlap is above some limit, then $\rho(q, \omega)$ is a
non-monotonic concave function of $q$, with a maximum at a given value
of $q$ in $[0,1]$ which depends on $\omega$.  Notice that, when
$\omega=1$ and $q=1$, i.e. if the two layers are identical and all the
nodes participate in both layers, the dynamics is identical to a voter
model on a single-layer network with $\mu_1 = \mu$. In fact, since all
the edges exist on both layers, a node participating in both layers
will have only $\mu$ distinct neighbours, and will be connected to
each of them on both layers. Hence, in a multiplex network with
$\omega=1$, the interface density $\rho$ of the voter model takes the
same value 0.22 at $q=0$ and at $q=1$, while for intermediate values
of $q$ the interface density is higher than that of a voter model on
each of the two layers.

We now consider the characteristic time $\tau$ of the multiplex
dynamics, where by characteristic time we understand twice the inverse
exponent of the multiplex survival probability $P^m_s(t)$.  We
consider the multiplex as active if at least one of the layers is
active, and find that for all $q > 0$ the survival probability of the
multiplex does decay exponentially with some exponent $\tau =
\tau(q,\omega)$ (fig.~\ref{fig:numericalresults:b}).  The only
exception is the $q = 0$ case of the fully-disconnected multiplex,
where the survival probabilities of the layers are independent.  In
this case the survival probability of the multiplex is given by the
probability that at least one of the layers is active, $P_s^m(t, q =
0, \omega = 0) \sim P_s(t)(2-P_s(t))$, or $P^m_s(t,0,0) \sim
2e^{-2t/\tau^{\rm single}} - e^{-4t/\tau^{\rm single}}$.  This means
that for $q = 0$ the multiplex survival probability \emph{does not
  scale exponentially}, and hence $\tau(q = 0)$ is not well-defined.
Fig.~\ref{fig:numericalresults:c} shows the characteristic times
$\tau(q, \omega)$ obtained (the value at $q = 0$ is that of an
approximate exponential fit).  The trend shows a peak at small $q$,
followed by an exponential decay with increasing number of interlayer
connections. The value of edge overlap $\omega$ controls the rate of
decrease.  We stress that the peak is \emph{not} a consequence of the
definition of $\tau$ for the multiplex, and the consequent rogue value
of $\tau$ at $q = 0$: the limit is truly singular, with plots for the
actual average time until absorption displaying the same features, as
do the plots for the characteristic exponent layer by layer. We also
note that the peak is robust with respect to system size. Therefore
the slowest finite-size multiplexes are ones where the layers are
interconnected by the smallest number of links. Multiplexes with more
interlayer connections will stop being active faster, as will wholly
disconnected systems. This insight can be understood by realizing that
in a multiplex with few interlayer links, the layers will most of the
time function as completely disconnected networks that half the time
will try to settle into the absorbing states of consensus different
from one another. Since this situation is now prohibited by the
few interlayer links, the multiplex will freeze only
when one of the layers switches over and both layers reach the same
consensus. It is this switching behaviour that is responsible for the
peak in $\tau$ for small $q$. We also notice that this non-linear
effect does not depend on the edge overlap $\omega$. Here, unlike with
the behaviour of interface density, the overlap does not change
qualitatively the behaviour of $\tau$ for increasing $q$. In fact, the
more interconnected the multiplex, the higher the role of edge
overlap: tuning up $\omega$ causes a decrease in the interface density
and results in a longer-lived multiplex, and the effect becomes more
pronounced with $q$.

\section{Irreducibility of the dynamics}
It was shown in \cite{Diakonova2014} that the interface density of the
multiplex voter model dynamics as a function of $q$
and in absence of edge overlap can be rewritten as
the interface density of a voter model on a single-layer network
having $\mu_1 = \avg{k}(1+q)$ (where $k$ is the average degree of each
of the two original layers), under some appropriately rescaled
time. Here it is important to note that the approximation of
\cite{Diakonova2014} treated the interlayer connections as inherently
probabilistic: the $q$ parameter was a probability that 
each node's state gets passed on to the other layer. It was shown that the
analytics of such a system, in the particular case $\omega=0$, are
equivalent to those of the voter model on some properly aggregated
single-layer network.  However, as Ref.~\cite{Diakonova2014} only
considered the thermodynamic limit, it is not obvious whether the
aggregate displays a corresponding rescaling of characteristic time.
Nevertheless, it should be possible to devise an aggregate that
results from the flattening of the multiplex into a single-layer
network, so that the resulting network will have a first moment of the
degree distribution equal to the expression $\mu_1$ given above.

The main question now is whether such a reducibility is possible in
the most general case in which the multiplex has edge overlap $\omega
\neq 0$. This is indeed the most interesting case, i.e.  when in a
multiplex there are correlations between the edges at the different
layers~\cite{Battiston2014,Nicosia2014corr}. In the following our
working hypothesis will be that, if the multiplex can indeed be
reduced to a monoplex (in the variables of interest), then there
exists an aggregate graph such that the behaviour of the voter model
on the multiplex is completely described by Eq.\eqref{eq:rho_single}
and Eq.\eqref{eq:tau_single} evaluated on the corresponding
monoplex. Since those equations depend just on the number of nodes and
on the first two moments of the degree distribution of the resulting
single-layer network, we will now
\begin{figure}%[!t]
	\centering
      \includegraphics[width=3.4in]{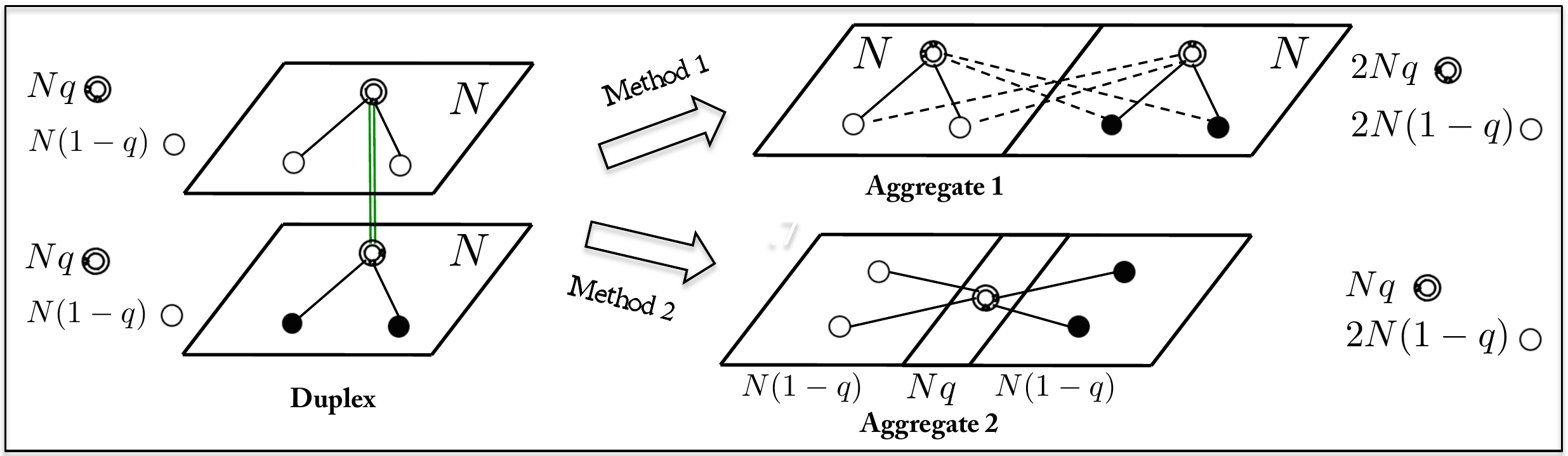}
  \caption{Schematic illustration of two possible aggregation
    procedures that reduce a two-layer multiplex with $N$ nodes on
    each to a single network.  The multiplex contains $qN$ interlayer
    links between associated nodes (pairwise links).  These nodes are
    shown in double circles. These nodes will have twice as many links
    in the aggregates, independent of the flattening method. Method 1
    keeps $2N$ nodes in the resultant aggregate, whereas method 2 has
    $N(2-q)$.}
  \label{fig:aggregation_schemas}
\end{figure}
consider two standard aggregation procedures, and derive analytically 
the values of the first and second moments of the resulting 
degree distributions. Let us remind that, by definition, the first 
moment of the degree distribution of a graph is equal to 
$  \mu_1 = \avg{k} = \sum_{k=1}^{\infty} k P(k) $, 
where $P(k)$ is the degree distribution, so that $P(k = \kappa)$
is the probability that the degree of a node sampled at random from
the graph is equal to $\kappa$. The probability $P(k)$ can be
also written as:
\begin{equation}
  P(k) = \frac{N_k}{\sum_{\ell}N_{\ell}}
\label{defpk}
\end{equation}
where $N_k$ is the number of nodes in the graph having degree equal to
$k$ and the normalization is just the total number of distinct nodes.

\begin{figure*}
	\hspace*{-0.2 in}
	\subfigure[Interface density $\rho$]{
    	\includegraphics[width=2in]{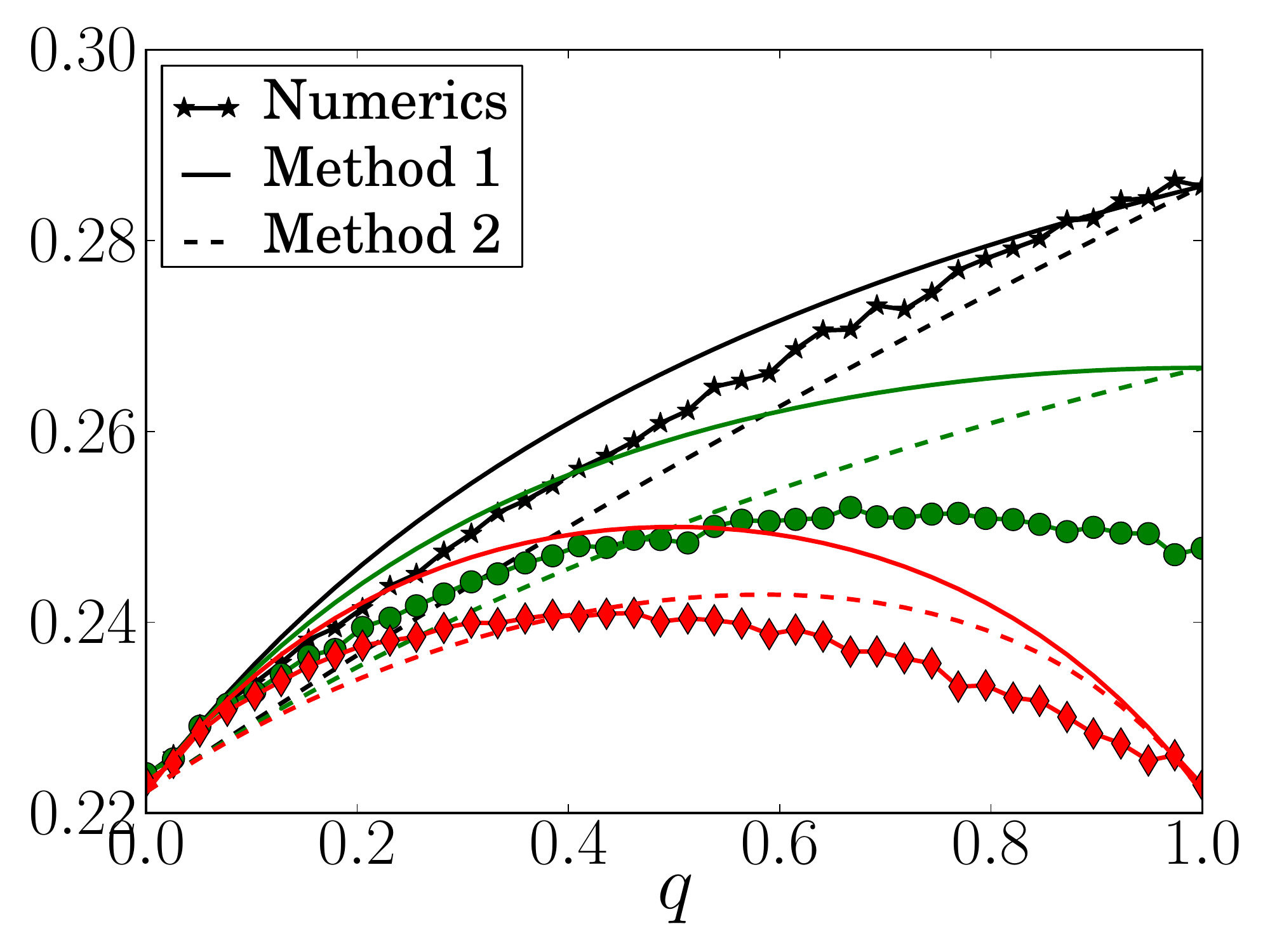}
		\label{fig:verification:a}
	}
	\hspace*{-0.2 in}
	\subfigure[Rescaled characteristic time $\tau/N$]{
    	\includegraphics[width=2in]{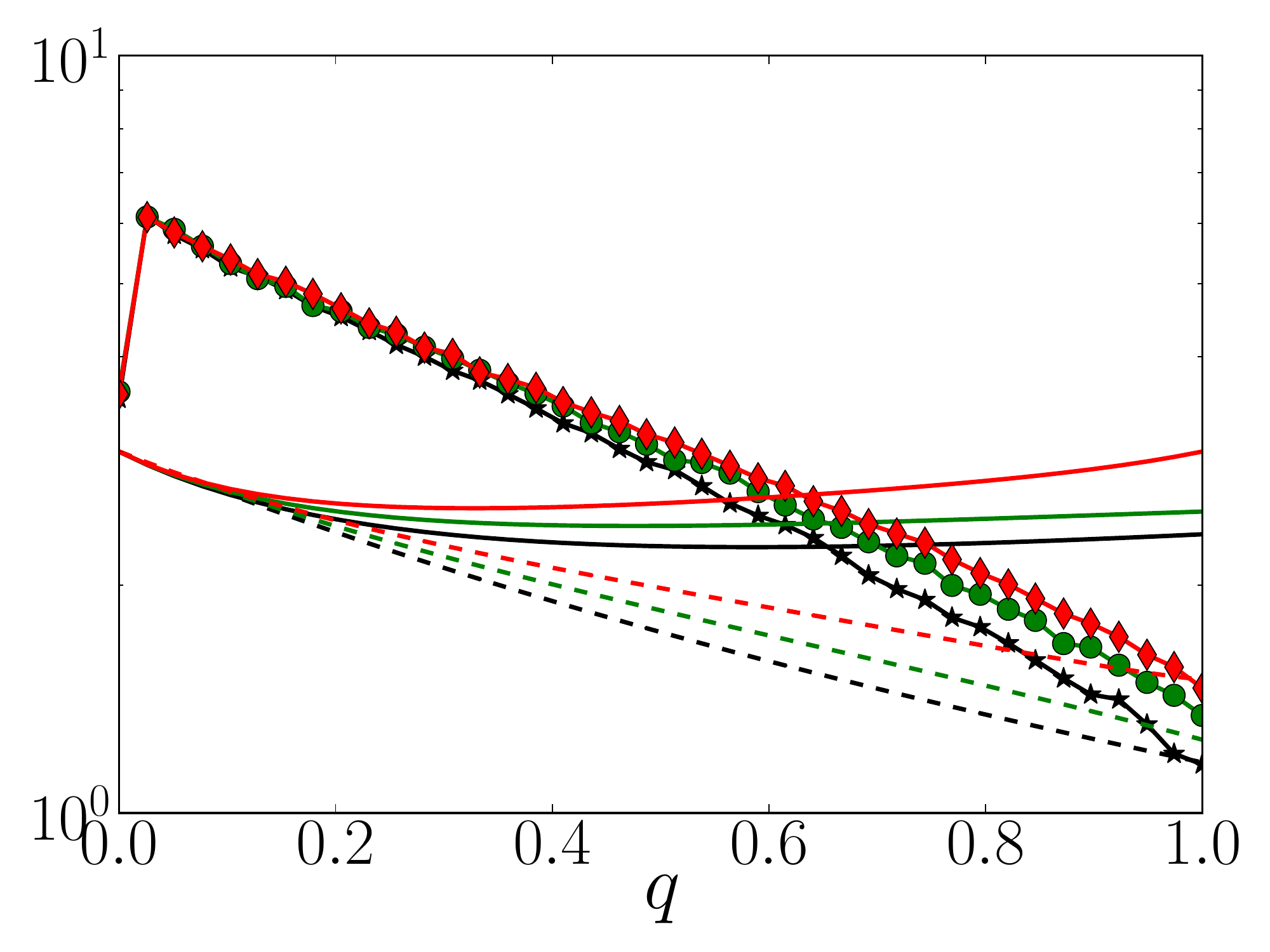}
        \label{fig:verification:b}
    }
	\hspace*{-0.2 in}
	\subfigure[Error of interface density]{
    	\includegraphics[width = 0.43\textwidth]{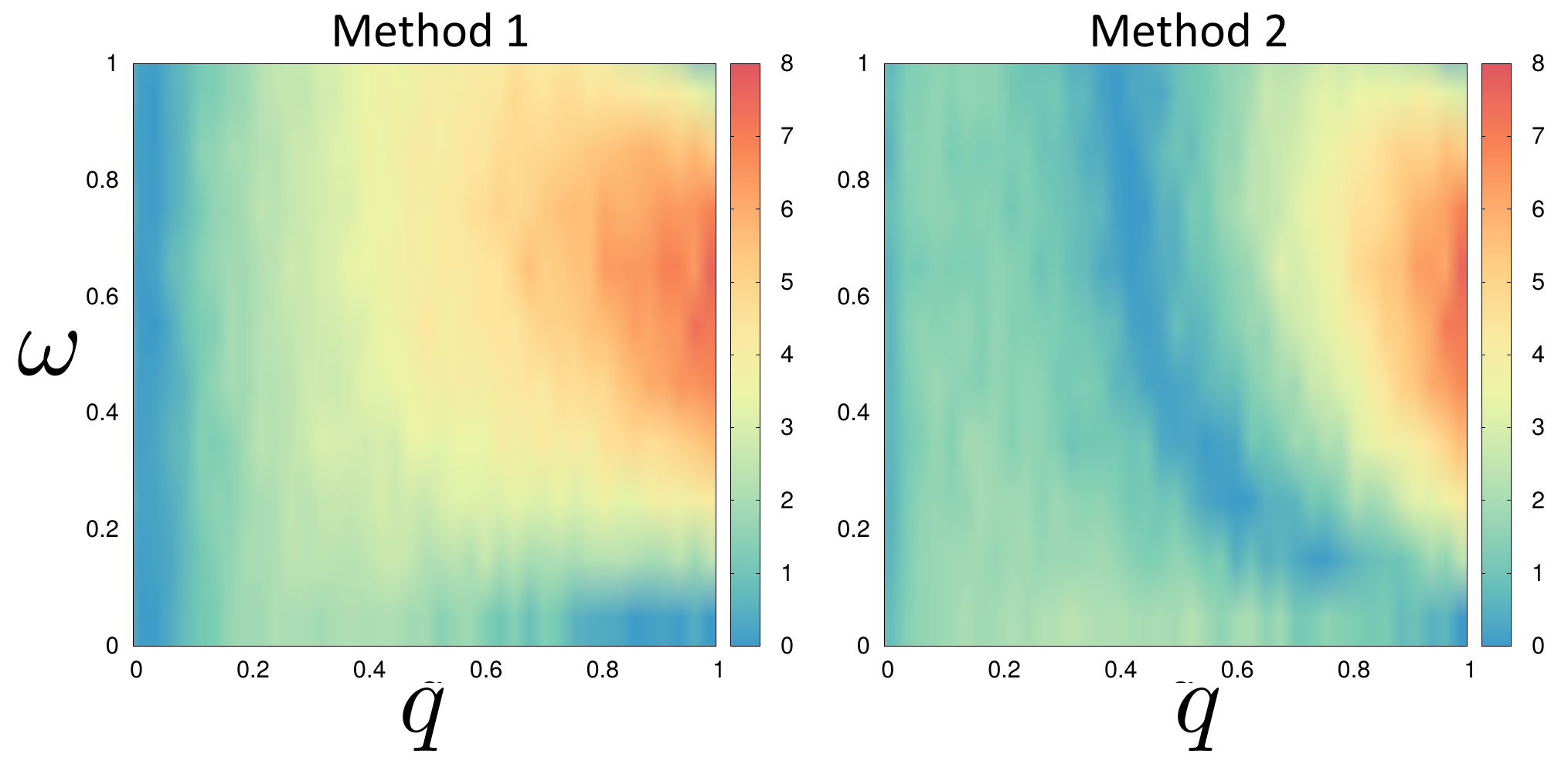}
		\label{fig:verification:c}
    }
  \caption{(Color online) Comparison of the theoretical prediction for
    (a) the interface density and (b) the characteristic time based on
    the aggregation of the duplex. Numerical results are shown for
    $\omega = 0$ (stars, black), $\omega = 0.5$ (circles, green) and
    $\omega = 1$ (diamonds, red), where the other parameters are the
    same as in Fig.\ref{fig:numericalresults}.  Analytics for the
    aggregates are computed for the respective values of $\omega$ for
    methods 1 and 2, and normalized by $N$. Panel (c): Absolute value
    of the relative error on the estimation of $\rho(q, \omega)$ using
    aggregation method 1 as a function of $q$ and $\omega$.}
  \label{fig:verification}
\end{figure*}

Given a duplex formed by two random regular graphs with identical
degree $\mu$, with edge overlap $\omega$ and where a fraction $q$ of
nodes participates in both layers, we can distinguish two classes of
nodes. The nodes in the first class exist in only one layer, and we
indicate as $k_{\rm single}$ the number of their neighbours, while nodes in the
second class exist in both layers (these are the nodes having an
inter-layer link) and we indicate their degree as $k_{\rm both}$. The
former class has degree:
\begin{equation} \label{k_single}
	k_{\rm single} = \mu,
\end{equation}
whereas nodes present on both layers have degree $k_{\rm both} = 2\mu$ 
when $\omega=0$.  
However, if the edge overlap is not null, i.e. when
$\omega > 0$, then the nodes being present in both layers 
have degree $k_{\rm both}$ equal to:
\begin{equation}\label{k_both}
  k_{\rm both} = 2\mu - q\mu\omega = \mu (2 - q\omega).
\end{equation}
In fact, the degree of a node $i$ present in both layers is equal to
the sum of its degrees on the two layers ($2\mu$) minus the expected
number of its edges which are present on both layers. This number is
equal to the probability that a neighbour $j$ of $i$ is also present
on both layers (which is equal to $q$), times the probability that the
edge $(i,j)$ is present in both layers (which is equal to $\omega$)
multiplied by the number of neighbours of $i$ (i.e., $\mu$). Hence we
get the correction $q\mu\omega$. In the particular case in which
$\omega=1$, we get $k_{\rm both} = \mu (2-q)$, while for $\omega=0$ we
recover $k_{\rm both} = 2\mu$.

Let us now consider the two following distinct aggregation procedures. 
They are illustrated in Fig.~\ref{fig:aggregation_schemas} and  
correspond to the two most standard ways to aggregate a duplex into 
a single-layer network. The two flattening procedures differ 
in the total number of nodes and also in the number of nodes of 
degree $k_{\rm single}$ and $k_{\rm both}$
that they produce. This in turn
changes the effective system size of the aggregate network, and the 
first two moments of the degree distribution.  In the following sections we
compute these quantities for the two methods, and assess how well the
aggregates fare in describing the behaviour of the true multiplex.

\begin{figure*}[!t]
  \begin{center}
    \subfigure[Effective degree distribution]{
	\hspace{-0.3in}
	\includegraphics[width=3.1in]{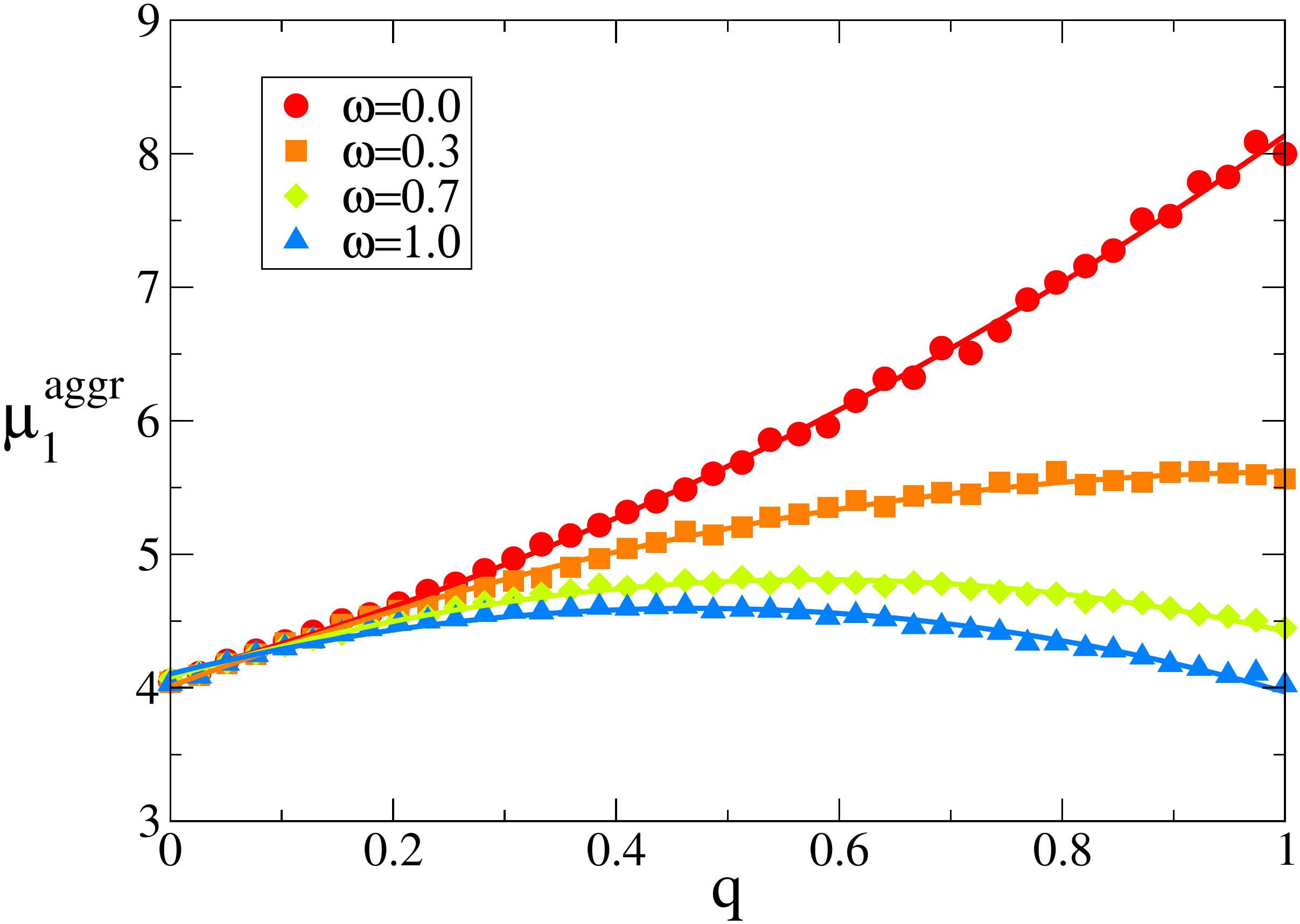}
	  \label{fig:nonlinear:a}
    }
    \subfigure[Scaling of characteristic time $\tau/N$]{
      \includegraphics[width=3.1in]{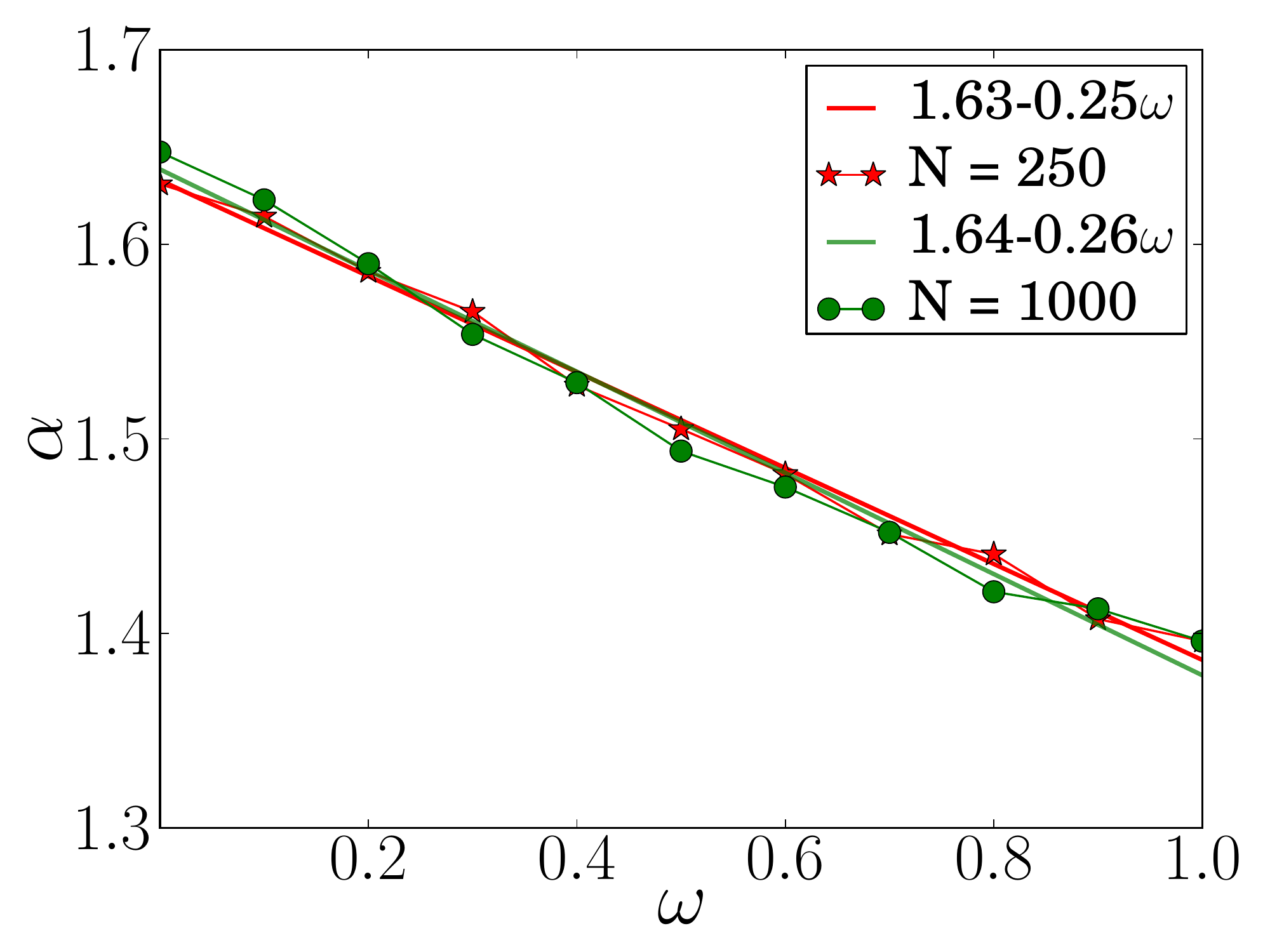}
	  \label{fig:nonlinear:b}
    }
  \end{center}
  \caption{(Color online) (a) The values of $\mu_1^{\rm aggr}$ resulting
    from the inversion of the measured interface density in the
    simulations as a function of $q$ for different values of edge
    overlap (symbols) and the corresponding quadratic fits
    (lines). (b) Scaling of $\tau$ with $q$ using the data shown in
    Fig.~\ref{fig:numericalresults}.  For each $\omega$, $\alpha$
    corresponds to the slope of $-ln(\tau/N)$ with $q$ computed for $q
    > 0$. Straight lines are the linear regression fits for each of
    the two $N$ trends, with coefficients given in the legend.}
  \label{fig:nonlinear}
\end{figure*}

\subsection{Aggregation method 1} % (fold)
\label{sub:method_1}
We can obtain a single-layer representation of a multiplex by putting
the layers side by side with no effective node overlap.  Therefore
$N_{\rm single} = 2N(1-q)$, $N_{\rm both} = 2Nq$, and the total number
of nodes in the aggregate is just $N_{\text{aggr}} = 2N$.  The degrees
of the nodes participating in just one or both layers are respectively 
equal to $k_{\rm single}$ and $k_{\rm both}$. 
By using Eq.~(\ref{defpk}) we have: 
\begin{equation}
	P(k=  k_{\rm single}  ) = (1-q),
\end{equation}
and
\begin{equation}
	P(k= k_{\rm both}  ) = q.
\end{equation}
The first moment ${\mu_1^{\rm aggr}}$ of the aggregated graph is then
equal to:
\begin{align}
  \displaystyle
 \mu_1^{\rm aggr} & = \sum_{k=1}^{\infty}k P(k) \nonumber\\
& = k_{\rm single} P(k= k_{\rm single}) +   k_{\rm both} P(k={k_{\rm both}})  
 \nonumber\\
& =  \mu_1 P(k={k_{\rm single}}) + \mu_1 (2 - q\omega) P(k={k_{\rm both}})  
 \nonumber\\
 	& = (1-q)\mu_1 + q(2\mu_1-q\omega \mu_1) \nonumber \\
	&= \mu_1(1 + q - \omega q^2),
\end{align}
and the second moment
\begin{align} \label{mu2_method1}
	\displaystyle
  	 \mu_2^{\rm aggr} & = (1-q)\mu_1^2 + q \mu_1^2 (2-q\omega)^2 \nonumber \\
	&= \mu_1^2 \left[ 1 + 3q -4 \omega q^2 + \omega^2 q^3 \right].
\end{align}
Under no overlap these reduce to
\begin{align} \label{mu_both_method1}
	\displaystyle
	& \mu_1^{\rm aggr} = \mu_1(1 + q), \nonumber \\
  	& \mu_2^{\rm aggr} = \mu_1^2 (1 + 3q).
\end{align}
Note that this method produces the same effective rescaling of the
first moment $\mu_1$ as given by the analytical estimation of the
interface density of the multiplex in the thermodynamic limit (see
Ref.~\cite{Diakonova2014}).

%=========================================================

\subsection{Aggregation method 2} % (fold)
\label{sub:method_2}
This method reduces the number of effective nodes
in the aggregate through treating nodes present
in both layers as \emph{one node}. So, while for
the rest of the nodes we still have
$N_{\rm single} = 2 N (1-q)$,
the number of `multiplex' nodes is now half as much
as in method 1, $N_{\rm both} = Nq$.\\
The total number of distinct nodes in the aggregate is equal
to $N_{\rm aggr} = 2N(1-q) + Nq = N(2-q)$. The degrees of
the single and `multiplex' nodes are as before
equal to $k_{\rm single}$ and $k_{\rm both}$ respectively. Therefore
\begin{equation}
	P(k= k_{\rm single}) = \frac{2(1-q)}{(2-q)},
\end{equation}
and
\begin{equation}
	P(k= k_{\rm both}) = \frac{q}{(2-q)}.
\end{equation}
Consequently, we have for the first moment of the degree distribution
$\mu_1^{\rm aggr}$: 
\begin{align}
  \displaystyle
\mu_1^{\rm aggr}
 = \sum_{k=1}^{\infty}k P(k) \displaystyle
 & =  \frac{2N (1-q)\times \mu_1 + Nq\times \mu_1 (2 - q\omega) }{N(2-q)} \nonumber\\
 \displaystyle
 & = \mu_1\frac{(2-q^2\omega)}{(2-q)}
\end{align}
and for the second moment $\mu_2^{\rm aggr}$:
\begin{align}
\mu_2^{\rm aggr} = \mu_1^2\frac{(q^3\omega^2 - 4q^2\omega + 2q +
    2)}{(2-q)}.
  \label{eq:m2_method2}
\end{align}
When the layers are uncorrelated and the overlap $\omega = 0$, we have
\begin{align*}
\mu_1^{\rm aggr} & = \frac{2\mu_1}{(2-q)}, \nonumber\\
\mu_2^{\rm aggr} & = 2\mu_1^{2}\frac{(1+q)}{(2-q)}
\end{align*}
We notice that, when $q=0$ ( i.e., if we have two non-interacting
layers), we get $\mu_1^{\rm aggr} = \mu_1$ and $\mu_2^{\rm aggr}
 = \mu_1^2 =
\mu_2$, while for $q=1$ we have $\mu_1^{\rm aggr}=2\mu_1$ and
$\mu_2^{\rm aggr} = 4\mu_1^2 = (2\mu_1)^2$.\\
%=========================================================

\subsection{Comparing aggregated to multiplex dynamics} % (fold)
\label{sub:assessing_the_methods}
We now compare the numerical results for the multiplex to the
theoretical values given by the corresponding equations for the
aggregates.  Eqns.~\eqref{eq:rho_single} and
\eqref{eq:tau_single} describe the
behaviour of the monoplex. We contrast the two aggregates by
substituting the respective effective value for the first and second
moments and the total number of nodes. In other words, we take $N
\rightarrow N_{\rm aggr}$, $\mu_1 \rightarrow \mu_1^{\rm aggr}$, and $\mu_2
\rightarrow \mu_2^{\rm aggr}$, where the effective values differ
depending on the aggregation method.

The results are reported in Fig.~\ref{fig:verification}. Consider
first the behaviour in the thermodynamic limit, described by the
interface density $\rho$. Both methods result in aggregates whose
qualitative behaviour with $\omega$ and $q$ corresponds to the trend
observed in the multiplex (Fig.~\ref{fig:verification:a}): for small
$q$ the system becomes more active with increasing number of
interlayer edges, whereas after a certain point activity may decrease,
depending on the whether or not the edge overlap is significant
enough. However, neither method gives correct quantitive understanding
for a general $\omega$ and $q > 0$.  This can be seen in
Fig.~\ref{fig:verification:c}, which compares the performance of the
two methods through computing the absolute value of the relative
error on $\rho$ as heat map of $q$ and $\omega$. Method 1 in general
does a better job than method 2 at being systematically consistent,
albeit suggesting higher values (the semicircular drop in the error to
zero, observed in method 2, comes about from the crossing of the
analytical and the numerical trends). As we have seen before, the
analytical results for Method 1 (at $\omega = 0$) correspond to a
multiplex with probabilistic interlayer connections of intensity
$q$. Therefore, at $\omega = 0$, comparing the numerical results for
the multiplex, and the analytical trend for method 1 can be used to
gain insight into the difference induced by an \emph{alternative
  method of inteconnecting the layers} of the multiplex (in fact, when
viewed in this way, the small magnitude of the differences becomes
more surprising than their presence).  The main quantitative
differences arise for a wide range of intermediate overlap values, for
medium-to-large interlayer connectivity - precisely the region of
parameter space motivated by real-world systems. The conclusion we
draw from this is that, as long as edge overlap is taken into account,
standard aggregates can only inform on some qualitative features of
large multiplex systems, but cannot capture quantitatively the
behavior of real multiplex networks.

Consider now the characteristic time of the approach to absorbing
states of finite-size systems (Fig.~\ref{fig:verification:b}). The
difference at $q = 0$ is due to $\tau(q = 0, \omega)$ not being defined for
the multiplex; yet the aggregate timescale is close. The monotonic
decrease for $q > 0$ shown by the numerics is captured only by method
2, with coincidence of $\tau$ values for $q = 1$. In fact, method 2
does a better job at both qualitative and quantitative results, unlike
method 1, which appears to work best in the thermodynamic 
limit (this aggregation
method results in a much slower system for large $q$ than the observed
duplex). The main discrepancy, however, is that neither of the methods
capture the jump of $\tau$ at small $q$. Therefore, although the two standard 
aggregation procedures can come close to describing the qualitative
features of long-term duplex activity, they are not sophisticated
enough to capture the long timescales associated with sparsely
interconnected systems. In the following Section we will see whether and 
how it is possible to devise more complex aggregation procedures in order 
to reproduce quantitatively the dynamics of the multiplex voter model.

\begin{figure*}
  \begin{center}
	\includegraphics[width=3in]{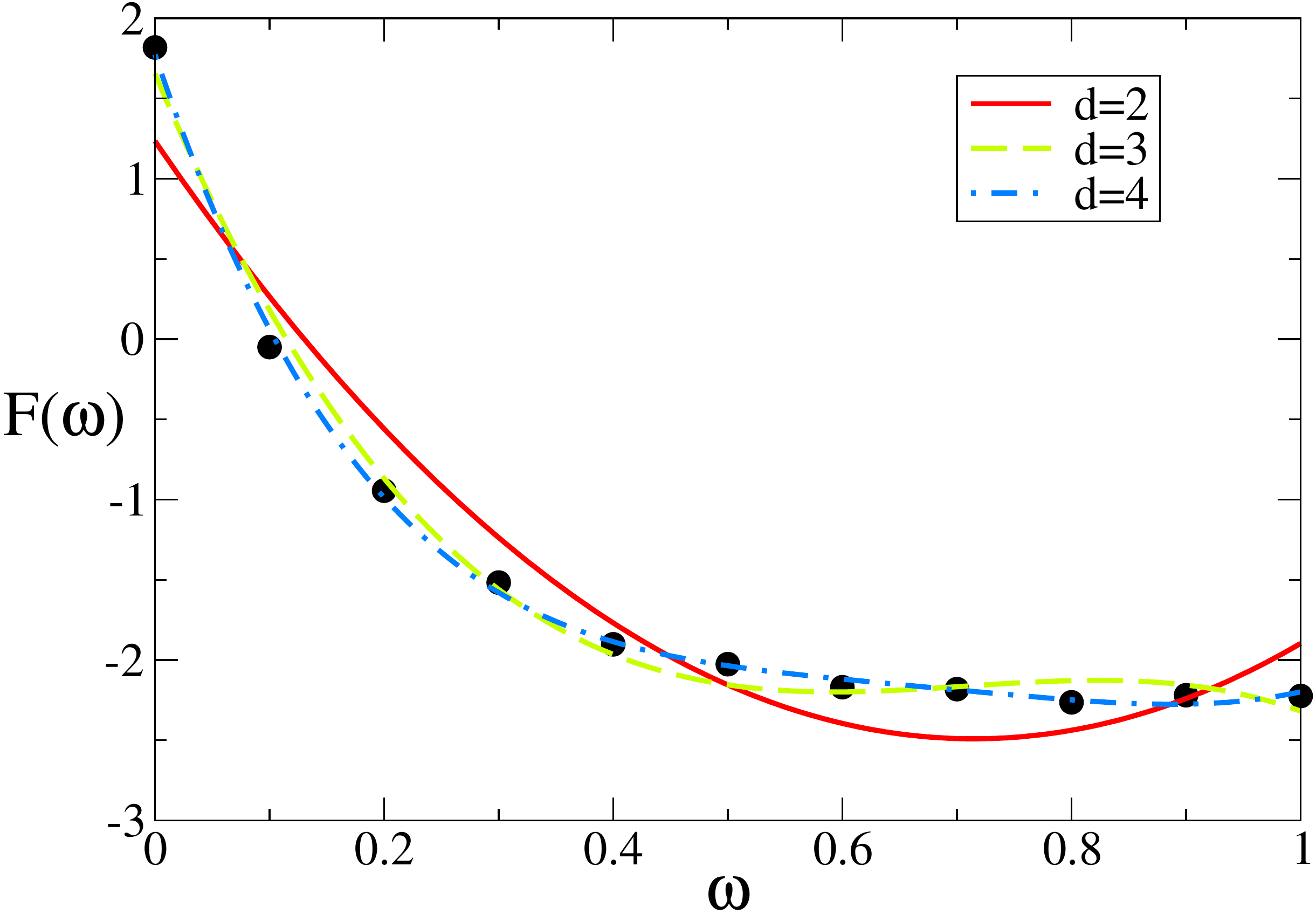}
	\includegraphics[width=3in]{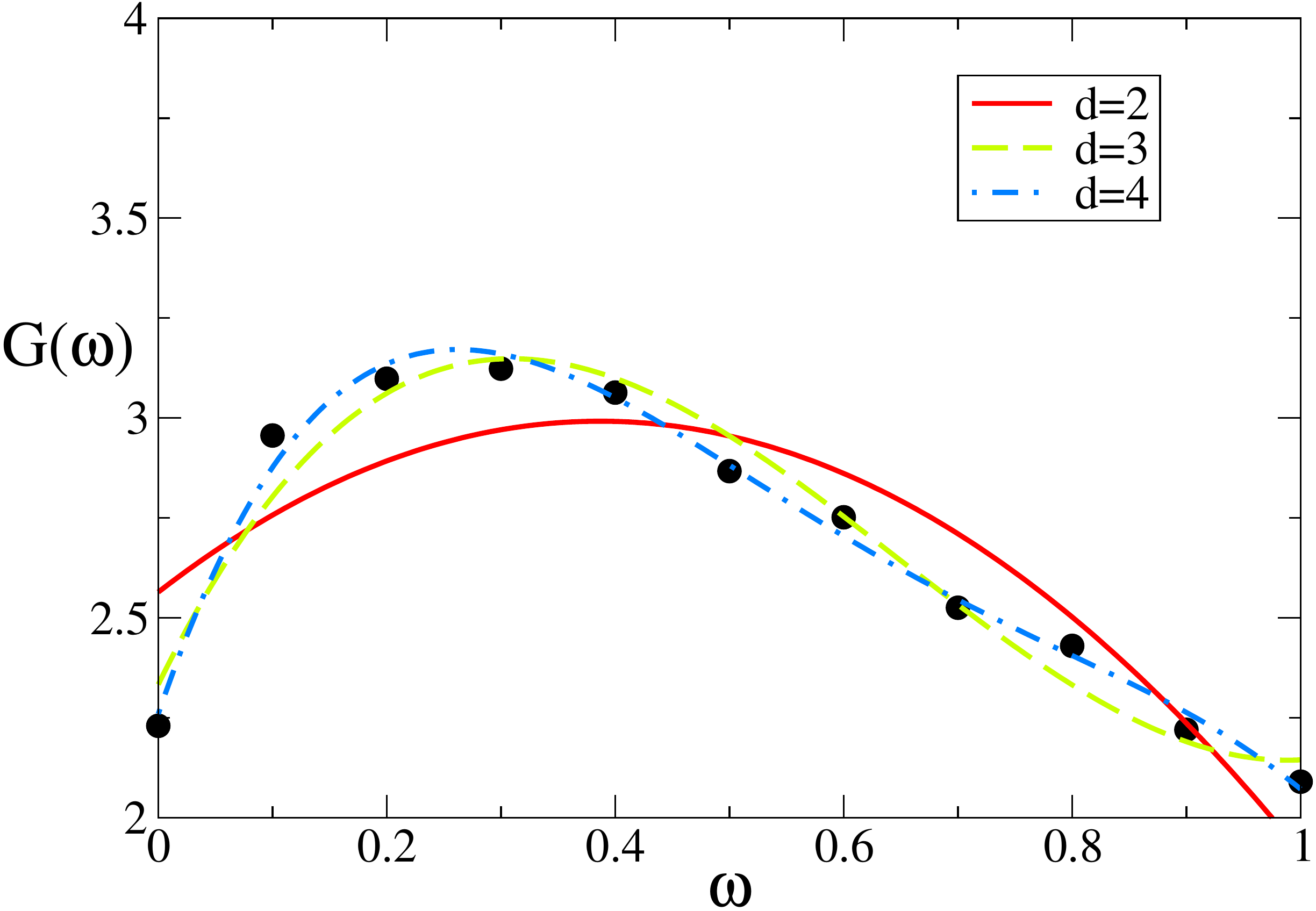}
  \end{center}
  \caption{The black dots in each panel represent the coefficients of
    the quadratic fit $F(\omega)q^{2} + G(\omega)q + 1$ of $\mu_1^{\rm
      aggr}(q,\omega)$ for different values of $\omega$. Notice that
    both $F(\omega)$ and $G(\omega)$ are non-linear functions of
    $\omega$. The different lines correspond to fits of $F(\omega)$
    and $G(\omega)$ using $d$-th order polynomials. Interestingly, a
    good fit is obtained only for $d\ge 3$, meaning that the presence
    of inter-layer state copying is introducing highly non-linear
    effects.}
  \label{fig:fit}
\end{figure*}

\section{Non-linear multiplex effects}

In principle, the results of the previous Section 
cannot absolutely rule out the possibility that there exist 
other aggregation methods reproducing 
the behaviour of $\rho(q, \omega)$ observed in the multiplex voter model.  
Our hypothesis is that the deviations from the theory found for
$\omega>0$ are due to the additional non-linearity induced by
inter-layer state copying made possible by the presence of a fraction
of inter-layer edges. In order to better investigate these non-linear
effects, we inverted Eq.~(\ref{eq:rho_single}) to compute the
effective value of the average degree $\mu_1^{\rm aggr}(q, \omega)$ of an
ideal aggregated single-layer network able to reproduce the observed value of
$\rho(q, \omega)$ for each value of $\omega$ and $q$. In formula:
\begin{equation}
  \mu_1^{\rm aggr}(q,\omega) = \frac{3\rho(q, \omega)
    -2}{3\rho(q, \omega) -1}
\end{equation}
As made evident by Fig.~\ref{fig:nonlinear}, for each value of edge
overlap, $\mu_1^{\rm aggr}(q,\omega)$ can be fitted very well by a
quadratic polynomial in $q$. However, the actual values of the
coefficients of the fit depend on $\omega$ in a non-trivial way. We
can formally write:
\begin{equation}
  \mu_1^{\rm aggr}(q,\omega) = F(\omega) q^2 + G(\omega) q + H(\omega).
\end{equation}
The problem now is to find an expression for $F(\omega)$, $G(\omega)$
and $H(\omega)$. We started by making an ansatz for the functional
dependence of those coefficients on $\omega$, and then fitting these
functions, for each value of $q$, by using several realisations of
$\mu_1^{\rm aggr}(q, \omega)$ corresponding to different values of
$\omega$.

We found that for all the values of $\omega$ the quadratic fit of
$\mu_1^{\rm aggr}(q,\omega)$ yields $H(\omega)=\mu$, so we just focused
on the other two coefficients. The values of $F(\omega)$ and
$G(\omega)$ are reported in the panels of Fig.~\ref{fig:fit} as black
circles. It is then reasonable to assume that $F(\omega)$, $G(\omega)$
are polynomial functions of $\omega$. We found that in order to
accurately reproduce the behaviour of $\mu_1^{\rm aggr}(q,\omega)$ in the
whole range of $\omega$, both $F(\omega)$ and $G(\omega)$ should be at
least third-order polynomials in $\omega$, as made evident by the
plots reported in Fig.~\ref{fig:fit}.

Notice that the predictions of $\mu_1^{\rm aggr}$ provided by the two
theoretical arguments reported above, based on the linear
superpositions of the two layers, contained only linear terms in
$\omega$. However, the fit of $\mu_1^{\rm aggr}(q, \omega)$ confirms
that, when $\omega>0$, the behaviour of the multiplex voter model is
the result of a highly non-linear combination of the two layers,
suggesting that a \textit{trivial} single-layer equivalent of the
multiplex voter model dynamics does not exist, especially when the
underlying multiplex network is characterised by a non-negligible
overlap.

That simple aggregation procedures do not produce the observed scaling
is also evident by examining the timescales on which finite systems
approach absorbing states. For $q > 0$, the characteristic time
$\tau(q, \omega)$ is an exponentially decreasing function of $q$,
$\tau(q, \omega)/N \sim e^{-q \alpha(\omega, N)}$, $\alpha > 0$
(fig.~\ref{fig:numericalresults:b}).  Figure~\ref{fig:nonlinear:b}
shows that $\alpha(\omega, N) = - a\omega + b$, $a > 0$, and that the
system-size dependency does not enter into it.  Hence,
$\tau(q, \omega)/N \sim e^{q(\omega a - b)}$.  Thus increasing the
edge overlap $\omega$ results in longer-lived systems, while adding
more interlayer links produces the opposite effect. This was
additionally confirmed by examining the behaviour of the rescaled time
until absorption $\left<T\right>$, which showed the same qualitative trend and almost
identical $a$ and $b$ coefficients.

\section{Discussion}

Multilayer networks allow to extend the applicability of network
theory to more realistic contexts in which nodes are connected through
concurrent interaction patterns of different kinds. However, a
fundamental open question to answer is whether the added complexity
yielded by multilayer networks is really needed to model network
phenomena, or if instead there exist simple ways of representing
multiplex dynamics through appropriately constructed processes
occurring on appropriately constructed single-layer networks. We have
investigated here the problem of reducing the multiplex voter model to
an equivalent single-layer dynamics. We have considered the
predictions about the level of activity of the multiplex voter model
in the thermodynamic limit, as measured by the interface density, as
well as the time to reach the absorbing state for finite systems. We
have found that results for the interface density based on
single-layer aggregated graphs are accurate only when there is little
or no interaction between the layers ($q$ sufficiently small), or when
there is full connectivity ($q \sim 1$) and the edge overlap $\omega$
is either $0$ or $1$.  For the complementary broad range of
parameters, any standard aggregation procedure can only give some
qualitative information about the interface density, but fails to
reproduce the quantitative details. We showed that edge overlap and
multiplexity have two opposite effects on the long-term dynamics of
the multiplex voter model, and in particular that an increase in the
value of edge overlap can counter the action of increasing the
fraction of interlayer links, leading to an overall \emph{decrease} in
the interface density of the multiplex. In fact, we showed numerically
that any equivalent single-layer representation of the multiplex voter
model dynamics entails the construction of an aggregate network which
is a highly non-linear combination of the original layers.

These results will be found at the same time surprising and
interesting by all the researchers aiming at modelling social
interaction in real-world scenarios. As a matter of fact, it has been
recently shown that multilayer social networks are normally
\textit{truly multiplex}, meaning that they are characterised by
non-negligible values of edge overlap~\cite{pewinternet} and by an
intermediate level of
multiplexity~\cite{Szell2010,Klimek2013,Nicosia2014corr}. And as we
have shown in this work, the single-layer voter model approximation is
qualitatively (when considering the characteristic time) and
quantitatively (when considering both the characteristic time and the
interface density) inaccurate when the edge overlap $\omega$ and the
degree of multiplexing $q$ are far from their extreme values $0$ or
$1$.  This is reflected in the average timescales of consensus for
finite systems, as we found that a multiplex with very few interlayer
connections takes the longest time to reach consensus, much more than
would probabilistically be needed by two disconnected monoplexes. This
nonlinear effect is not captured by either of the simple aggregation
procedures proposed in the paper.

The case of dynamical irreducibility of a multiplex process presented
in this work raises the important question of whether other unknown
phenomena might be lurking in the multilayer structure of real-world
systems. This question, together with the insights about the
intrinsically multidimensional nature of the multiplex voter model,
represents a stimulus to perform further research along these lines.

\begin{acknowledgments}
  This work has been supported by the Spanish MINECO and FEDER under
  projects INTENSE@COSYP (FIS2012-30634), and by the EU Commission
  through the project LASAGNE (FP7-ICT-318132).
\end{acknowledgments}

\end{document}